\documentclass[a4paper,12pt]{article}

\usepackage[english]{babel}
\usepackage{xcolor}

\usepackage{array} 

\usepackage{authblk}

\usepackage[a4paper,top=2cm,bottom=2cm,left=2cm,right=2cm,marginparwidth=1.75cm]{geometry}

\usepackage[backend=biber,style=numeric-comp, sorting=none]{biblatex}
\usepackage{setspace}
\usepackage[left]{lineno}

\usepackage{amsmath}
\usepackage{graphicx}
\usepackage{csquotes} 
\usepackage[colorlinks=true, allcolors=blue]{hyperref}

\title{Steerable dual-trap optical tweezers with confocal position detection using back-scattered light}

\author{Md Arsalan Ashraf}
\author{Pramod Pullarkat}
\affil{Raman Research Institute, Bangalore -- 560080, India}

\date{}

\begin{document}
\begin{spacing}{1.2}
    \maketitle
\end{spacing}

\begin{abstract}
\begin{spacing}{1.2}
Optical tweezers has emerged as a powerful tool in manipulating microscopic particles and in measuring weak forces of the order of a pico-Newton. As a result, it has found wide applications ranging from material science to biology. Dual-trap optical tweezers (DTOT) are of particular importance as they allow for two point correlation measurements as in molecular force spectroscopy, two-point active micro-rheology, etc. Here we report a novel design for a steerable DTOT setup which uses back-scattered light from the two traps for position detection. This is performed using a confocal scheme where the two detectors are placed at the conjugate points to the respective traps. This offers several significant advantages over current designs, such as, zero cross-talk between signals, single module assembly and robustness to thermal drift.  Moreover, our design can be very easily integrated with standard microscopy techniques like Phase contrast and Differential Interference Contrast, without modifying the microscope illumination unit.

\end{spacing}
\end{abstract}

\section{Introduction}
Optical tweezers was introduced by Sir Arthur Ashkin in 1970 \cite{ashkinAccelerationTrappingParticles1970, ashkinObservationSinglebeamGradient1986, ashkinForcesSinglebeamGradient1992}, a discovery that earned him the Nobel prize in 2018. It is a contact-free or non-invasive technique where small dielectric particles, including living cells, or even single atoms can be trapped using gradient forces generated by light. The ability of optical tweezers to measure extremely weak forces at the microscopic scale--of the order of thermal fluctuations \cite{huangDirectObservationFull2011, gomez-solanoExperimentalVerificationModified2009}, single molecule forces \cite{bustamanteOpticalTweezersSinglemolecule2021, zaltronOpticalTweezersSinglemolecule2020, capitanioInterrogatingBiologyForce2013}, and forces generated during the stepping of single molecular motor proteins \cite{svobodaDirectObservationKinesin1993, sudhakarGermaniumNanospheresUltraresolution2021} and its application in micro-rheology \cite{addasMicrorheologySolutionsSemiflexible2004, atakhorramiShortTimeInertialResponse2005} studies has led to an explosion in its application in physics, material science, and life sciences. Even after more than four decades since its initial applications, optical tweezers are being constantly improved leading to new discoveries \cite{sudhakarGermaniumNanospheresUltraresolution2021, bustamanteOpticalTweezersSinglemolecule2021}. At present there are several designs for single, dual and multiple optical traps that are in use for a variety of applications \cite{volpeRoadmapOpticalTweezers2023, bustamanteOpticalTweezersSinglemolecule2021}. Of these, dual-trap optical tweezers (DTOT) are particularly interesting because of their applications in measurement of microscopic interaction potentials \cite{crockerMicroscopicMeasurementPair1994, meinersDirectMeasurementHydrodynamic1999}, optical binding studies \cite{burnsOpticalBinding1989}, two point micro-rheology \cite{starrsOneTwopointMicrorheology2003, buchananComparingMacrorheologyOne2005, paulDirectVerificationFluctuationdissipation2017, paulTwopointActiveMicrorheology2018}, single molecule force measurement \cite{bustamanteOpticalTweezersSinglemolecule2021, meinersFemtonewtonForceSpectroscopy2000, sungSingleMoleculeDualBeamOptical2010, meijeringNonlinearMechanicsHuman2022}, etc.

 Dual trap constructions typically uses two separate laser beams that are derived either by splitting a single laser into two using appropriate optics or by using two separate lasers \cite{fallmanDesignFullySteerable1997, atakhorramiTwinOpticalTraps2008, paulTwopointActiveMicrorheology2018}. To steer the traps in lateral direction (in the focal plane of the microscope objective) the incoming beams are tilted about the back focal point of the microscope objective. Position/force measurements are usually performed by collecting the light scattered in the forward direction by the trapped particles. In this method, a position detector, usually a Quadrant Photodiode (QPD), is employed to detect any shift in the position of the trapped particle with respect to the trap centre via the interference of the direct and scattered light. One of the disadvantages of this method of detection is the presence of cross-talk between the signals emanating from the two traps \cite{atakhorramiTwinOpticalTraps2008}. Avoiding/minimizing this requires far more complex optical designs or use or separate lasers with different wavelengths, which significantly increases the cost of design. Another disadvantage of using forwards scatter detection schemes, especially when implemented on advanced research microscopes, is its need for specialized optics that replaces the microscope illumination system. This prevents some microscopy techniques like Phase Contrast imaging and Differential Interference Contrast imaging, often used for soft matter or biological samples, and limits the possibility of implementing micro-manipulation techniques like use of micro-needles, or integration of optical trapping with Atomic Force Microscopy, magnetic tweezers, etc. Apart from these limitations, the use of forward scattered light for detection becomes challenging when used for force measurements or to characterize rheological properties of highly heterogeneous materials, for example tissue, where wavefront distortions cause errors in detection \cite{favre-bulleOpticalTrappingVivo2019}. Another situation where transmitted light detection becomes an issue is when the sample is highly absorptive or has opaque interfaces. 
 
 To avoid or minimize the above mentioned limitations, position/force detection can be performed using back-scattered light, by collecting this signal using the same microscope objective that is used for trapping. However, in current steerable DTOT designs, this becomes cumbersome as the back-scattered light from each trap, when made to focus on to a detector, shifts in position as the trap position is shifted around \cite{paulTwopointActiveMicrorheology2018}. This means that the detection optics need to be realigned each time either one or both traps are moved. Moreover, cross-talk is not eliminated in this method as detection is performed using polarization splitting and back focal plane interferometry \cite{paulTwopointActiveMicrorheology2018}. In this article we present a novel design for a dual-trap optical tweezers where we employ a confocal detection scheme for the back-scattered light used for position/force detection. In this scheme, position detection occurs at the conjugate image plane of the microscope and cross-talk is completely eliminated by the use of pinholes. This allows for position/force detection with high spatial and temporal resolution simultaneously in both traps, and also ensures that position/force detection can be performed continuously as the two traps are steered independent of each other. Furthermore, we demonstrate that this design is robust to several sources of drift arising from thermal expansions of components.

\section{Optical design}
    
\begin{figure*}[tbhp!]
    \centering
    \includegraphics[width=\textwidth]{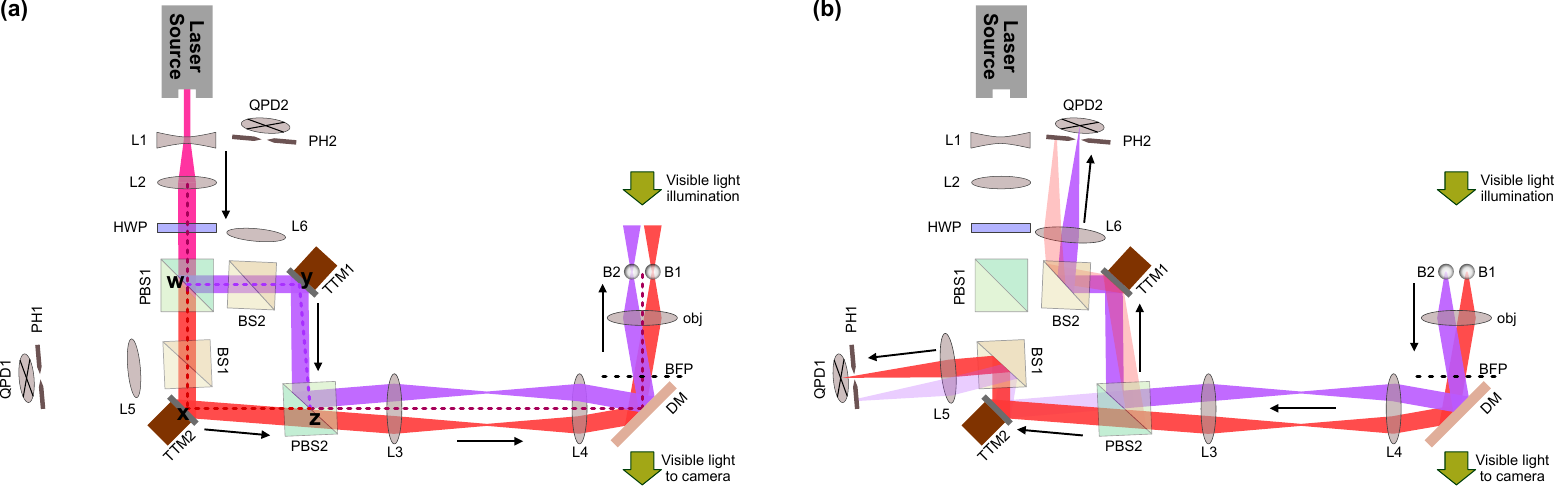}
    \caption{\footnotesize Schematic diagrams showing the optical designs used for trapping and detecting particles in our DTOT which uses detection of back-scattered light in a confocal arrangement. (a) Optics and beam paths used to create an independently steerable DTOT. The abbreviations for optical components are explained in Table \ref{ListofComponents}. Black arrows indicate the direction of propagation of light beams. Horizontally polarized light is shown in red and vertically polarized light is shown in purple. The darker dashed lines show the beam path for creating either trap at optic axis of the objective. (b) Beam paths for independent detection the position of the trapped particles using back-scattered light. Black arrows indicate the direction of propagation of light beams. Red beam is the light coming from the first trapped bead B1, and purple beam is the light coming from the second trapped bead B2.}
    \label{schematic}
\end{figure*}

\begin{table}
\caption{}
\begin{center}
\begin{tabular}{m{0.12\textwidth}|m{0.38\textwidth}}
     \hline
     \hline
     component & details  \\
     \hline
     \hline
     Laser & 3W continuous wave laser ($\lambda$ = 1064 nm), opus 1064, Laser Quantum \\
     \hline
     L1 & achromatic doublet (f = -50 mm), ACN254-050-B, Thorlabs\\
     \hline
     L2 & achromatic doublet (f = 100 mm), AC254-100-C-ML, Thorlabs\\
     \hline
     L3 & plano-convex lens (f = 250 mm), LA1301-B, Thorlabs\\
     \hline
     L4 & achromatic doublet (f = 500 mm), AC508-500-C-ML, Thorlabs\\
     \hline
     L5, L6 & achromatic doublet (f = 250 mm), AC254-250-C-ML, Thorlabs\\
     \hline
     HWP & half wave plate ($\lambda$ = 1064 nm), WPH10ME-1064, Thorlabs\\
     \hline
     PBS & polarizing cube beamsplitter, CM1-PBS253, Thorlabs\\
     \hline
     BS & non-polarizing cube beamsplitter, 10:90 (R:T), BS044, Thorlabs\\
     \hline
     TTM & Piezo tip/tilt mirror, S-335, Physik Instrumente\\
     \hline
     DM & short pass dichroic mirror, cutoff wavelength 903 nm, ZT1064rdc-sp, Chroma Technology\\
     \hline
     obj & 100X (1.3 NA) oil immersion objective, Fluorescent/Semiapochromat, UPLFLN100XO2, Olympus\\
     \hline
     QPD & quadrant photodiode, SPOT-9DMI, OSI Optoelectronics\\
     \hline
\end{tabular} 
\end{center}
\label{ListofComponents}
\end{table}

We describe our dual-trap optical tweezers in two parts: the creation of the traps and the position detection.   

\subsection{Creating two trapping points} 

A schematic diagram of the setup is shown in fig. \ref{schematic}a (see Table \ref{ListofComponents} for specifications and catalog numbers of parts used). For the construction of the traps, a diode laser emitting continuous wave of polarized TEM00 mode at 1064 nm wavelength and a maximum power of 3 W is used as the light source. The diameter of the laser beam is 1.8 mm which is expanded using a combination of a concave lens (L1) and a convex lens (L2) having focal lengths of -50 mm and 100 mm, respectively. The expanded beam then passes through a half wave plate (HWP) mounted on a rotary mount which enables the selection of a linear polarization state. A polarizing beam splitter (PBS1) then splits the beam into two orthogonally polarized beams: the transmitted beam, shown in red, is horizontally polarized and the reflected beam, shown in purple, is vertically polarized. The red and purple beams then pass through non polarizing beam splitters BS1 and BS2 respectively (R:T = 10:90), the role of these beam splitters is explained in next section. The two beams are then reflected from tip/tilt mirrors (TTM1 \& TTM2), which are used to independently steer the two traps. These mirrors are mounted such that the reflection surfaces undergo pure tilt (the beams reflect from fixed points on each reflecting surfaces). After reflection from the tip tilt mirrors, the two beams enter another polarizing beam splitter (PBS2). The horizontally polarized beam (red) is transmitted while the vertically polarized beam (purple) is reflected by PBS2. Both the beams then goes through a pair of convex lenses (L3 \& L4) having focal length 250 mm and 500 mm respectively. The distance between L3 and L4 is equal to sum of their focal lengths, so that it makes a afocal system and the beam coming out from L4 remains collimated.
The beams are then reflected into the microscope objective (obj) using a short pass dichroic mirror (DM). The objective lens focuses the beams at its front focal plane to form two optical traps where dielectric particles with refractive index higher than the surrounding medium can be trapped. Note that L3 and L4 are positioned in such a way that the back focal point of L3 coincides with the reflecting surfaces of TTM1 \& TTM2, and the front focal point of L4 lies on the back focal plane (BFP) of the objective lens (obj). In this configuration, the back focal plane of the objective and reflecting surface of either of the two tip tilt mirrors makes a conjugate pair. Using this arrangement, the two traps can be independently moved around in the front focal plane of the objective by titling the computer controlled tip tilt mirrors. Steering is performed either using a double joystick gaming console or using a home-developed LabView software.

Important thing to note is that the centre of PBS1, the centres of the reflecting surface of tip tilt mirrors, and the centre of PBS2 (marked as WXYZ in fig. \ref{schematic}a) are placed on the four corners of a right trapezoid, the sides xz and yz are of same length. This is done to ensure two things (i) the lens L3 is equidistant from the two tip tilt mirrors (as both of these mirror should lie on back focal plane of L3), and (ii) the surfaces of PBS2 are not lying perpendicular to the propagating beam, this ensures that any reflection from these surfaces,which could potentially cause error in position detection, do not retrace the beam path. An example of reflection from the surfaces, when the beam splitter is lying perpendicular to the laser beam, is shown in fig. \ref{FilteringSpots}a. The unwanted light spots disappears (fig. \ref{FilteringSpots}b) when the beam splitters are mounted at an oblique angle.

\subsection{Position detection using back-scattered light}

\begin{figure}[tbhp!]
    \centering
    \includegraphics[width=0.5\textwidth]{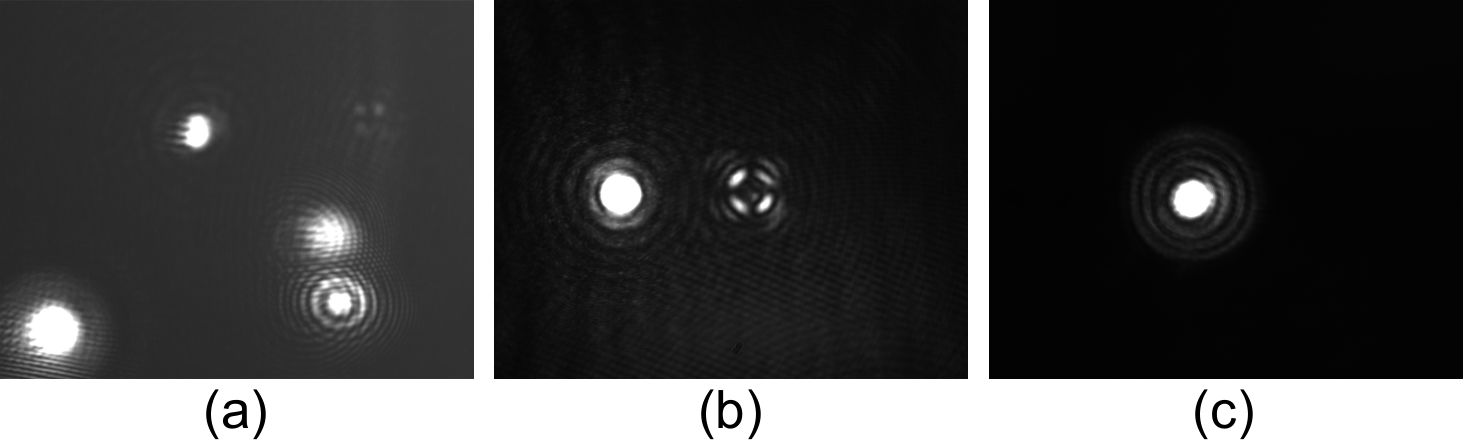}
    \caption{\footnotesize (a) Intensity pattern that form at the detector plane due to the back-scattered light from the two trapped beads and unwanted reflections arising from the beam splitters. The pattern is imaged using a camera at the detector plane of one of the detectors.   (b) The unwanted reflections are removed by placing the beam splitters at oblique angle, retaining only the back-scattered light from the two beads. (c) To detect the position of the first bead, the spot from the second bead is blocked by placing a pinhole. Note that when the two traps are independently steered, the back-scattered light from the first bead will remain centred on the pinhole, allowing for its detection by the QPD.}
    \label{FilteringSpots}
\end{figure}

A schematic diagram showing the detection optics and the paths of the back-scattered light from trapped particles (B1 and B2) is shown in fig. \ref{schematic}b. The trapped particles (usually micron sized, spherical, dielectric beads) have refractive index higher than the surrounding medium and hence they scatter the laser light with which the traps are created. The light scattered in backward direction is collected by the objective lens (obj) and reflected towards the lens pair L4, L3 by the dichroic mirror DM. Majority of light from the trapped bead B1 (shown in red) gets transmitted through the polarizing beam splitter PBS2 as the scattered light retains the polarization of the trapping beam to some extent, but at the same time, a small fraction of this light gets reflected from PBS2 (shown in faint red). Similarly majority of light from the trapped bead B2 is reflected by PBS2 (shown in purple), but a small fraction of it gets transmitted (shown in faint purple). 

The light transmitted by PBS2 (arising from the two beads, shown in red and faint purple) is reflected towards the non-polarizing beam splitter BS1 (reflection to transmission ratio 10:90). A majority of these light beams get transmitted through BS1 (not shown in the figure) and a small fraction is reflected towards the lens L5 (f = 250 mm), which focuses the two beams on to the quadrant photodiode (QPD1). The QPD1 is aligned with the optic axis of L5 and placed one focal length away from L5. In this configuration the trap position for the bead B1 and the centre of QPD1 make a confocal pair. Because of this, the back-scattered light from B1 get focused at the centre of QPD1 while the light coming from B2 is focused away from the centre of the QPD1 as shown in fig. \ref{schematic}b. In order to detect the position of B1, independent of B2, the back-scattered light from B2 is blocked using a pinhole PH1 (1mm diameter). Images taken using a camera at the position of the QPD, with and without the pinhole are shown in  fig. \ref{FilteringSpots}b \& c, respectively. Similarly, reflected light from PBS2 (purple and faint red) arrives at BS2 (reflection to transmission ratio 10:90) and a small amount of these beams are reflected towards the lens L6 (f = 250 mm) which focuses them onto QPD2. Centre of QPD2 and the position of the second trap makes a conjugate pair, and thus the back-scattered light from B2 falls at the centre of QPD2. Light from B1 falls away from the centre and is blocked by the pinhole PH2. Now if the first bead moves relative to the first trap position (due to an external force), then the movement is detected by QPD1, and similarly movement of the second bead relative to second trap position is detected by QPD2. This detection scheme is akin to a confocal microscope, where the light from the focal point is collected by a pinhole placed at a conjugate point to allow point by point scanning. As the back-scattered light from the two beads traces independent paths, and are finally separated by the pinholes using a confocal setting, cross-talk is completely eliminated. This is one major advantage of this detection scheme over detection using forward-scattered light.

The diameter of the pinholes is 1 mm, and is chosen such that it is larger than the image of the trapped bead. This allows all the light from the bead to pass through even if the bead moves away from the trap point. The pinhole size should be large enough to capture bead displacement at-least in the linear regime of the QPD, but should be small enough to cut the light from the bead trapped in the second trap. With a 100X objective our setup gives a magnification factor of $\approx$ 300X  at the QPD, a 1 \textmu m sized bead will make 300 \textmu m sized image at the QPD, which means for 1mm sized pinhole, the inter-trap distance should be more than 2.2 \textmu m (approximately) for independent detection. If two traps come any closer than that, the light from both trapped particles will enter the two QPDs resulting in unwanted cross-talk. 

\subsection{Imaging of the trapped bead and sample}
\begin{figure*}[tbh]
    \centering
    \includegraphics[width=\textwidth]{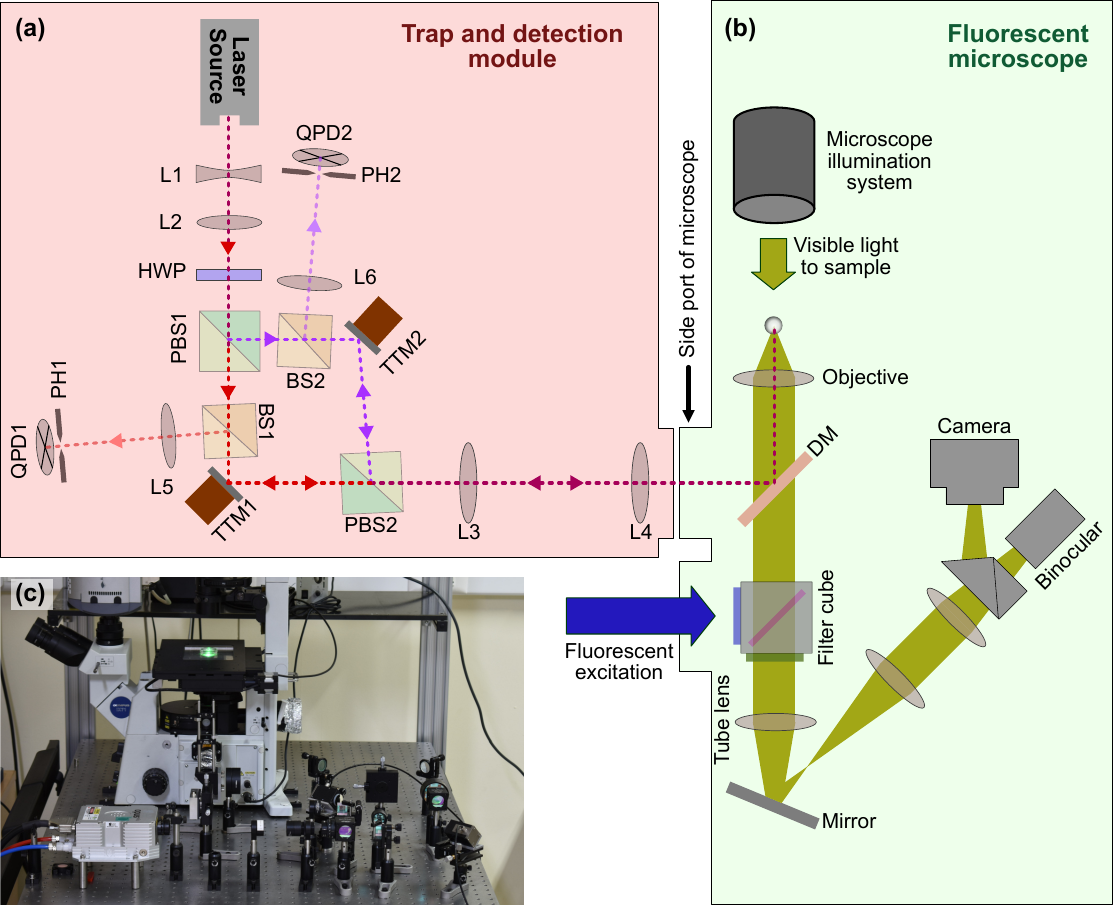}
    \caption{\footnotesize (a,b) Schematic showing how the optical trap module is coupled to the fluorescence microscope. (c) Image of the DTOT setup built around an Olympus-IX71 microscope. All the optics for optical tweezers, except IR dichroic mirror (DM), are mounted outside the microscope. The dichroic mirror (not visible in the image) is mounted below microscope objective. The condenser of the microscope is untouched.}
    \label{FullSchematic}
\end{figure*}

Since the dichroic mirror DM (short pass, cutoff wavelength 903 nm) transmits wavelengths below that of the infrared laser, imaging of the trapped objects or any other sample can be performed using visible light sources (see fig. \ref{FullSchematic}). Moreover, as the condenser part of the commercial microscope (Olympus IX71) is left untouched, this arrangement allows for phase-contrast microscopy without any alterations to the microscope. As an infinity corrected objective is used, Differential Interference Contrast (DIC) microscopy can also be implemented easily with only a slight modification--the prism slider has to be introduced after the dichroic mirror DM. The dichroic mirror is placed above the fluorescence turret of the microscope and hence the system allows for fluorescence microscopy in the visible range as well. These three imaging modes are widely used in material science studies, especially when using biological samples.

\section{Characterization and testing of the setup}

\subsection{Detector calibration}

\begin{figure}[tbhp!]
    \centering
    \includegraphics[width=0.5\textwidth]{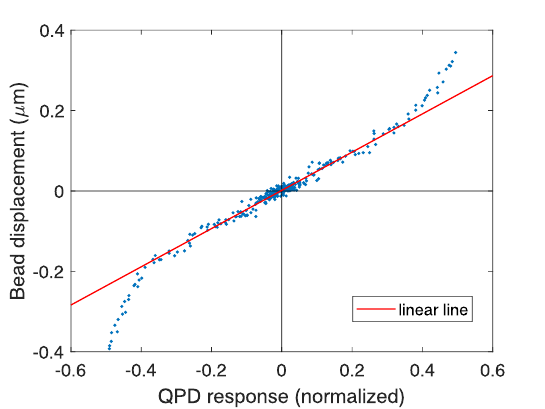}
    \caption{\footnotesize QPD response obtained by displacing a trapped bead (2.9 $\mu$m) using fluid drag force. A gravity driven flow chamber setup was used to generate laminar flow. The displacement of the trapped bead was obtained using camera based tracking. As can be seen by comparing with the straight line, the QPD detection is linear upto 200 nm.} 
    \label{QPDCalibration}
\end{figure}

For high frequency position detection of the trapped particle, we have used QPDs with 10 $\mu$m gap (SPOT-9DMI, OSI Optoelectronics) in our setup (fig. \ref{schematic} \& \ref{FullSchematic}). The QPD was attached to a sum and difference amplifier (QD50-0-SD, OSI Optoelectronics), which is a transimpedance amplifier that gives three output signals--one sum and two difference signals. In order to obtain a higher gain factor, we have modified this commercial amplifier board by replacing the feedback resistors.  The native gain of the amplifier was $10^4$, which was  increased to $47\times10^4$ by this modification.

We used the drag force method to calibrate the QPD response. For this, we trapped a bead inside a flow cell and a transient flow was introduced to impose a drag force on the bead, shifting it from the trap centre. The changes in bead position was imaged using a CCD camera and the actual displacement was determined using a video tracking algorithm \cite{cheezumQuantitativeComparisonAlgorithms2001}. The QPD response for a 2.9 $\mu$m sized bead is shown in fig. \ref{QPDCalibration}. The response is linear for displacements up to 200 nm on either side.

\subsection{Checking for cross-talk}
\begin{figure}[tbhp!]
    \centering
    \includegraphics[width=0.5\textwidth]{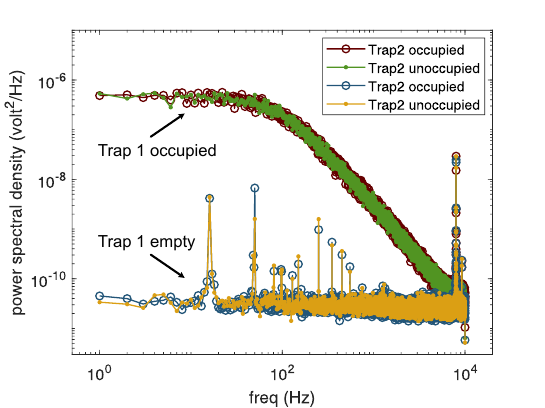}
    \caption{\footnotesize Power spectral density (PSD) of displacement signal from QPD1 under four different cases. (case i) Plot with yellow dots is obtained when both traps are empty. (case ii) Plot with red hollow circles when both traps are occupied. (case iii) Green data when trap 1 is occupied while trap 2 is empty. (case iv) Blue circle when trap 1 is empty while trap 2 is occupied. PSDs from trapped bead or noise floor remain unaffected by the presence/absence of bead in another trap, this shows that there is no cross-talk between the two detectors and independent measurement from two traps are possible.
    } 
    \label{CrossTalk}
\end{figure}
In order to check for possible cross-talk between the detected signals, we recorded the signals under the following conditions.  (i) Both traps were on but no bead were trapped in either of them; (ii) Both traps were occupied by beads; (iii) Only trap 1 was occupied; and (iv) only trap 2 was occupied. We record the signal from trap 1 and calculates its power spectral densities while changing the occupancies of the two traps. When both traps are unoccupied (case i), we obtain the power spectrum of the noise floor (yellow curve in fig. \ref{CrossTalk}). When both traps are occupied (case ii), one obtains a Lorentzian curve as expected (red circles in fig. \ref{CrossTalk}) \cite{berg-sorensenPowerSpectrumAnalysis2004}. When trap 1 is occupied and trap 2 is unoccupied (case iii), one obtains a Lorentzian shown by the green curve. For the last case, \emph{ie.} trap 1 is unoccupied and trap 2 is occupied, one detects only the detector noise floor (blue circles). 
Had there been any cross-talk between the two detections, the QPD1 would have detected the dynamic signal from the bead in trap 2 and the PSD would have looked different from the noise floor recorded in absence of bead in trap 2. This is a major advantage of the current design over detection schemes that use polarization based dual trap arrangement combined with forward-scattered detection scheme.

\subsection{Robustness against thermal drift in the setup}
\begin{figure}[tbhp!]
    \centering
    \includegraphics[width=0.5\textwidth]{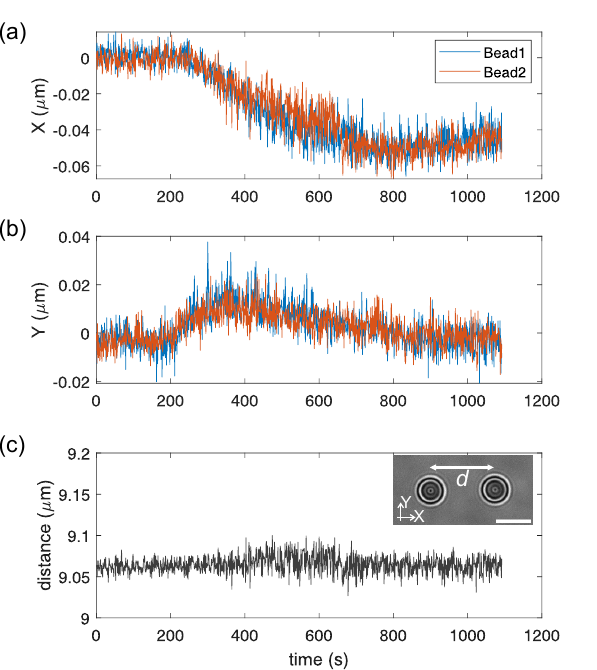}
    \caption{\footnotesize Plots showing long time ($\approx$ 18 min) position measurement of two trapped particles (shown in the inset of (c)) without any applied force. The position detection was done using camera tracking as the QPD signal won't register the drift. (a,b) X \& Y positions of the two trapped particles relative to their respective initial position. The beads show similar drift in X or Y direction indicating both of the trap position is moving together with respect to camera. This is seen in plot (c) which shows the distance $d$ between the centres of the beads.} 
    \label{DriftData}
\end{figure}

Temperature changes may alter the optical alignments which can cause the trap positions to drift. In our setup, as the two traps are built using same laser source and most of the optics are common between the two beams, any drift that are caused by common optical elements placed after PBS2 will move both the trap positions in same direction. Hence, there is no relative displacement of the two traps even though individual trap positions may exhibit drift. In our system, the QPDs only detect the relative positions of the trapped particles irrespective of the trap position and hence the setup is immune to drift when measuring forces of objects held between the two traps. In order to test this, we have characterized the drift in our setup by using the camera to detect the positions of the trapped particles. Fig. \ref{DriftData}(a \& b) shows the X \& Y position time series of both the particles relative to the their respective initial positions. Fig. \ref{DriftData}(c) shows the the centre to centre distance between the two particles. As can be seen, the distance between the two trapped particles remains the same on the average over long intervals of time.  

\section{Discussion and comparison with other dual optical trap techniques}

\begin{figure}[tbhp!]
    \centering
    \includegraphics[width=0.5\textwidth]{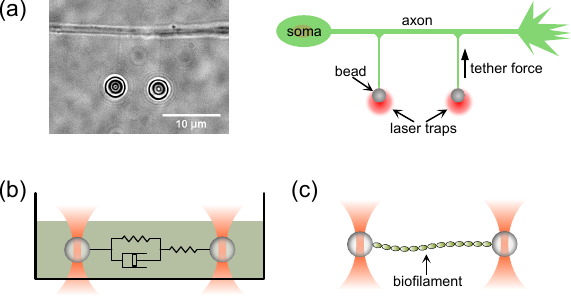}
    \caption{\footnotesize Various possible experiments that can be performed using our DTOT setup. (a) An image and a schematic of two membrane nanotube pulled from an axon of a neuronal cell using our DTOT in order to study biomechanics of cell surface. A schematic of this scheme is shown on the right. (b) Schematic for two point active micro-rheology to study viscoelastic properties of soft material like polymer gels. (c) A bio-filament tethered between two trapped beads to perform single molecule force spectroscopy.} 
    \label{ExampleExp}
\end{figure}

Several schemes have been used in the past for dual-trap optical tweezers \cite{fallmanDesignFullySteerable1997, moffittDifferentialDetectionDual2006, atakhorramiTwinOpticalTraps2008, visscherConstructionMultiplebeamOptical1996, meinersDirectMeasurementHydrodynamic1999, bustamanteHighResolutionDualTrapOptical2009, pesceOpticalTweezersTheory2020, finerSingleMyosinMolecule1994, sungSingleMoleculeDualBeamOptical2010}. The most common method is to use a polarizing beam splitter to obtain two orthogonally polarized beams which are used for trapping as in our case. However, the detection is done using the forward scattered light \cite{moffittDifferentialDetectionDual2006, meinersDirectMeasurementHydrodynamic1999, bustamanteHighResolutionDualTrapOptical2009, sungSingleMoleculeDualBeamOptical2010}. This requires an additional optical module that replaces the regular condenser of the microscope and this module has to be maintained in perfect alignment with the trapping module. Another major disadvantage of this technique is that the separation of the forward scattered light arising from the two traps based on the polarization state of the respective trapping beams is not perfect. This is because the forward scattered light undergo partial depolarization and hence cross-talk is inevitable. Some designs overcome the latter problem by using two lasers of different wavelengths \cite{atakhorramiTwinOpticalTraps2008}, but this has a cost disadvantage, apart from making techniques like fluorescence microscopy more challenging. 

We have demonstrated a novel design for steerable dual-trap optical tweezers, in which position detection is done using back-scattered light from the trapped particles. A confocal design scheme is used for detection where the detector centre is conjugated with the trapping point. This way, the image of the trap remains centred on the QPD detector even when the trap is being steered and position (force) detection becomes possible while trap is being moved. 
A major advantage of this confocal scheme is that the force measurement is immune to drift in the system. This is because (i) the image of the trap centre always remain centred on the QPD even when the trap drift and (ii) because the distance between the two trap centres remain constant even when the individual traps drift. The latter advantage is critical when performing single molecule force spectroscopy according to the scheme shown in fig. \ref{ExampleExp}c. A second major advantage is that cross-talk between the signals arising from the two traps is completely eliminated. This is because the image of the second trap point falls away from the centre of the detector belonging to first trap and is cut off by the pinhole. A third major advantage is that as our design uses only back-scattered light, the whole trapping setup along with the position detection can be assembled as a single module and integrated to a research microscope via a side port (see fig. \ref{FullSchematic}. This also means that the illumination path of the research microscope (condenser) remains intact allowing for various imaging modes like Phase contrast, Differential Interference contrast, etc \cite{murphyFundamentalsLightMicroscopy2012}. Moreover, the back-scatter detection scheme, in general, also allow for integration of the optical traps with other micro-manipulation or force measuring techniques like AFM, micro-needles, etc. Apart from the fact that the region above the sample is free for integrating these techniques, this scheme also has the advantage that only the lower part fo the sample need to be transparent and optically flat \cite{keyserOpticalTweezersForce2006}.

Although the technique we have presented here has several advantages, it also presents some challenges and disadvantages as well. Some of these can be overcome easily as follows. One of the main disadvantage is the low intensity of the back-scattered light and loss of further intensity upon reflections from the beam splitters. However, this can be mitigated by adopting the following methods. (i) the amplification factor of the QPDs can be enhanced by using custom electronics as in our case--commercial sum/difference circuits are designed to operate with much higher light intensities. (ii) The splitting ratio of the beam splitters can be altered to favor the back propagating beams while maintaining the intensity of the trapping beams using a higher laser power. For most force measurement studies a 3W laser provides ample intensity for the two beams. Another minor disadvantage is that there are weak reflections arising from the coverslip-water interface when the trap position is close to the coverslip. We observe that this is almost completely eliminated if the trap is 10\textmu m or more away from the glass-water interface. 

Since its discovery almost fifty years back, optical trapping has seen constant development in both design and applications \cite{volpeRoadmapOpticalTweezers2023}. Today, it is a major quantitative tool in material science research and especially so in quantitative biophysical studies. We believe that this new design, with its multiple major advantages that we have detailed above, will contribute significantly to both basic science and applied research. 

\section{Acknowledgement}
We thank Meena M S, Sandhya and Raghunathan A for help with electronics, Md Ibrahim for machining mechanical parts, and Nishant Joshi for fruitful discussions on optics design. The authors acknowledge support through The Wellcome Trust DBT India Alliance (grant IA/TSG/20/1/600137).

\begin{spacing}{1.2}
    \printbibliography
\end{spacing}

\end{document}